\documentclass{article}
\usepackage{dcase2022,amsmath,url,times,booktabs,tabularx,amsmath,amssymb}
\usepackage{array,multirow}
\usepackage{arydshln}
\usepackage{multicol}
\usepackage{footnote}
\usepackage{float}
\usepackage{cite}
\usepackage{enumitem}
\usepackage{color,soul}

\usepackage[pdftex]{graphicx}

\newcolumntype{C}[1]{>{\centering\arraybackslash}p{#1}}
\newcolumntype{L}[1]{>{\raggedright\arraybackslash}p{#1}}
\newcolumntype{R}[1]{>{\raggedleft\arraybackslash}p{#1}}

\title{MIMII DG: Sound Dataset for malfunctioning industrial machine investigation and inspection for domain generalization task}

\name{Kota Dohi,
Tomoya Nishida,
Harsh Purohit,
Ryo Tanabe,
Takashi Endo}
\secondlinename{
Masaaki Yamamoto,
Yuki Nikaido, 
and Yohei Kawaguchi}

\address{Research and Development Group, Hitachi, Ltd.\\
1-280, Higashi-koigakubo, Kokubunji, Tokyo 185-8601, Japan\\
\texttt{\{kota.dohi.gr, yohei.kawaguchi.xk\}@hitachi.com}}

\begin{document}

\ninept
\maketitle

\begin{sloppy}

	\begin{abstract}
		We present a machine sound dataset to benchmark domain generalization techniques for
		anomalous sound detection (ASD).
		Domain shifts are differences in data distributions that can degrade the detection performance, and handling them is a major issue for the application of ASD systems.
		While currently available datasets for ASD tasks assume that occurrences of domain shifts are known, in practice, they can be difficult to detect.
		To handle such domain shifts, domain generalization techniques that perform well regardless of the domains should be investigated.
		In this paper, we present the first ASD dataset for the domain generalization techniques,
		called MIMII DG.
		The dataset consists of five machine types and three domain shift scenarios for each machine type.
		The dataset is dedicated to the domain generalization task with features such as multiple different values for parameters that cause domain shifts and introduction of domain shifts that can be difficult to detect, such as shifts in the background noise.
		Experimental results using two baseline systems
		indicate that the dataset reproduces domain shift scenarios and is useful for benchmarking domain generalization techniques.
	\end{abstract}

	\section{Introduction}
	\label{sec:intro}
	Anomalous sound detection systems (ASD) are automatic inspection systems that identify
	anomalous sounds emitted from machines
	\cite{koizumi2017neyman, kawaguchi2017how, koizumi2019neyman,kawaguchi2019anomaly,
		koizumi2019batch, suefusa2020anomalous, purohit2020deep, dohi2021flow}.
	Because these systems use microphones to conduct inspections,
	contactless inspections of anomalies inside the machines can be realized, unlike
	the vibration monitoring systems \cite{carden2004vibration, toh2020review, Khademi2019}.


	For the widespread application of ASD systems, researchers have mainly tackled two types of challenges.
	First, in real-world cases, only a few anomalous samples are available or provided anomalous samples
	do not cover all possible types of anomalies. Therefore, unsupervised anomaly detection methods are
	often adopted so that the system can detect anomalies by training with only normal samples.
	MIMII \cite{purohit2019mimii} and ToyADMOS \cite{koizumi2019toyadmos} are the first datasets that contain machine sounds in
	real factory environments, and are used for benchmarking the performance of
	unsupervised ASD methods.

	Second, the detection performance of the system degrades due to changes in the distribution of normal sounds (i.e., domain shifts).
	Domain shifts for an ASD task can be classified into two categories;
	operational domain shifts caused by changes in states of a machine and environmental domain shifts
	caused by changes in the background noise or in the recording environment.
	One solution for handling domain shifts is to use domain adaptation techniques and adapt the model to the new data.
	MIMII DUE \cite{tanabe2021mimiidue} and ToyADMOS2 \cite{harada2021toyadmos2} were developed for benchmarking domain adaptation
	techniques, while an unsupervised scenario was also assumed.

	However, in some real-world cases, domain generalization techniques \cite{zhou2021domain,Wang2020,Dissanayake2021} rather than domain adaptation techniques can be preferred. For example, if the operational domain shifts occur too frequently, adaptation of the model can be difficult. This is because only a small amount of data can be used for adaptation and frequent adaptation can be too costly. For another example, if domain shifts are difficult to detect, such as the domain shifts in the background noise, adaptation of the model can also be difficult. In these cases, domain generalization can be useful for handling domain shifts. Because these techniques aim at generalizing the model to detect anomalies regardless of the domains, adaptation of the model during the operation is not necessary.  Therefore, domain generalization techniques for ASD task should be investigated for handling domain shifts that are too frequent to adapt or too difficult to detect.


	To benchmark the domain generalization techniques for ASD task, a new dataset dedicated to the domain generalization task should be developed.
	This is because the data required for domain generalization and domain adaptation can be different. For example, generalization of the model may require a larger number of sets of data recorded under different conditions.
	Also, because domain generalization techniques are likely to be used for domain shifts that can be difficult to detect, this type of shifts should be included in the dataset for domain generalization tasks.

	In this paper, we present a new dataset for benchmarking ASD methods
	using domain generalization techniques. The dataset consists of five different machine types;
	fan, gearbox, bearing, slide rail, and valve. Each machine type includes
	three sections, each of which corresponds to a type of domain shift.
	Each section consists of the source domain data to be used for generalizing the model and the target domain data
	for evaluating the domain generalization performance.
	The source domain has at least two different sets of values that cause domain shifts to generalize the model. Also, domain shifts that can only be handled with domain generalization techniques are included in the dataset.
	The dataset is freely available at \url{https://zenodo.org/record/6529888}
	and is a subset of the dataset for Task 2 of the DCASE 2022 Challenge.

	\section{Recording environment and setup}
	\label{sec:typestyle}
	We prepared five types of machines (fan, gearbox, bearing, slide rail, and valve), three types of factory noise data (factory noise A, B, and C), and three different domain shift scenarios for each machine type.
	The types of machines and domain shift scenarios were chosen on the basis of our experiences building ASD systems for real-world commercial solutions. Here, we identify each scenario of domain shifts by {\bf section IDs}.
	The details of the type of domain shift for each section and the values of the parameters that shift between domains, the domain shift parameters, are described in Table~\ref{parameters}.

	We then recorded sound data of each machine to reproduce the domain shift scenarios we assumed. We recorded both normal and anomalous sounds for each domain, where to reproduce anomalous sounds, we used deliberately damaged machines or operated machines in an incorrect manner. For recording, we used a TAMAGO-03 microphone manufactured by \textit{System In Frontier Inc.} \cite{tamago}. The recording was conducted either in a sound-proof room~(Fan and Valve) or in an anechoic chamber~(Gearbox, Bearing, Slide rail).
	Although the microphone has eight channels, we only used the first channel for the dataset.
	Recorded sound clips are 16-bit audio with a sampling rate of 16 kHz and are 10~seconds long.
	Examples of spectrograms for each machine type are shown in Figure 1.
	A short description and recording procedures of each machine type are as follows.\\

	\noindent
	\textbf{Fan} An industrial fan used to keep gas or air flowing in a factory.
	Operational conditions were kept the same between source and target domains, since Fan was dedicated to environmental domain shifts.
	Anomaly types include wing damage, unbalanced, clogging, and over voltage.\\

	\noindent
	\textbf{Gearbox} A gearbox that links a direct current (DC) motor to a slider-crank mechanism, transmitting the power generated by the rotation of the motor at a constant speed to the slider-crank mechanism. The slider-crank mechanism then converts the rotational motion into a linear motion and raises and lowers its weight. We changed the operation voltage and mass of the weight to cause domain shifts. Anomaly types include gear damage and over voltage. \\

	\noindent
	\textbf{Bearing} Two ball-type bearings are attached to a shaft with a spindle motor, and the sound is emitted from the bearing as it supports the rotating shaft. We changed the rotation speed of the shaft and the location of the microphones to cause domain shifts.
	Anomaly types include eccentricity in the bearing for two different directions.\\

	\noindent
	\textbf{Slide rail (slider)}
	A linear slide system consisting of a moving platform and a staging base that repeats a pre-programmed operation pattern. We changed the operation velocity and acceleration to cause domain shifts. Anomaly types include cracks on the rail, removal of grease, and a loose belt for a belt-type slide rail.\\

	\noindent
	\textbf{Valve}
	A solenoid valve that repeatedly opens and closes in accordance with a pre-programmed operating pattern and is connected to a pump to control air or water flow. We changed the operating pattern and location of the panels surrounding the valve. Anomaly types include contamination in the valve.\\

	\begin{table*}[t]
		\begin{center}
			\caption{Type of domain shift, values of domain shift parameter, and SNR for each section. Values of domain shift parameters represent machines the sound of which are mixed in Fan section 00, levels of noise in Fan section 02, and locations of microphone in Bearing section 01.}
			\label{parameters}
			\begin{tabular}{@{}l l r l l l@{}}
				\hline
				\multicolumn{2}{c}{\begin{tabular}{@{\hskip0pt}l@{\hskip0pt}} \textbf{Machine type / }\\ \textbf{section ID} \end{tabular}} &
				\begin{tabular}{c@{\hskip0pt}c@{\hskip0pt}} \textbf{SNR [dB]} \end{tabular}                     &
				\begin{tabular}{c@{\hskip0pt}c@{\hskip0pt}} \textbf{Type of Domain shift}\\ \textbf{[Domain shift parameter]}  \end{tabular}                     &
				\begin{tabular}{c@{\hskip0pt}c@{\hskip0pt}} \textbf{Parameter values for }\\ \textbf{source-domain}  \end{tabular}                     &
				\begin{tabular}{c@{\hskip0pt}c@{\hskip0pt}} \textbf{Parameter values for}\\ \textbf{target-domain}  \end{tabular}
				\\
				\hline
				\multirow{6}{*}{Fan}
				                                              & 00
				                                              & -6.0
				                                              & \multirow{2}{14em}{Mixing of different machine sound [machine sound index]}
				                                              & W, X
				                                              & Y, Z
				\\\\
				                                              & 01
				                                              & -12.0
				                                              & \multirow{2}{14em}{Mixing of different factory noise [factory noise index]}
				                                              & A, B
				                                              & C
				\\\\
				                                              & 02
				                                              & N/A
				                                              & \multirow{2}{14em}{Different levels of noise [noise level~(SNR~[dB])]}
				                                              & L1~(3), L2~(-9)
				                                              & L3~(-3), L4~(-15)
				\\\\
				\hline
				\multirow{6}{*}{Gearbox}
				                                              & 00
				                                              & -6.0
				                                              & Different operation voltage [V]
				                                              & 1.0, 1.5, 2.0, 2.5, 3.0
				                                              & \multirow{2}{12em}{0.6, 0.8, 1.3, 1.8, 2.3, 2.3, 3.3, 3.5}
				\\\\
				                                              & 01
				                                              & -12.0
				                                              & \multirow{2}{14em}{Different weight attached to the gearbox [g]}
				                                              & 0, 50, 100, 150, 200                                                                          & 30, 80, 130, 180, 230, 250
				\\\\
				                                              & 02
				                                              & -12.0
				                                              & Different gearbox ID [machine ID]
				                                              & 05, 08, 13
				                                              & 00, 02, 11
				\\\\ \hline
				\multirow{5}{*}{Bearing}
				                                              & 00
				                                              & 12.0
				                                              & Different rotation speed [krpm]
				                                              & 6, 10, 14, 18, 22                                                                             & 2, 4, 8, 12, 16, 20, 24, 26
				\\
				                                              & 01
				                                              & 12.0
				                                              & \multirow{2}{14em}{Different microphone location [location of the mic.]}
				                                              & A, B, C, D                                                                                    & E, F, G, H
				\\\\
				                                              & 02
				                                              & 12.0
				                                              & \multirow{2}{14em}{Mixing of different factory noise [factory noise index]}
				                                              & A, B
				                                              & C
				\\\\ \hline
				\multirow{6}{*}{Slide rail}
				                                              & 00
				                                              & -6.0
				                                              & Different operation velocity [mm/s]
				                                              & 300, 500, 700, 900, 1100
				                                              & \multirow{2}{12em}{100, 200, 400, 600, 800, 1000, 1200, 1300}
				\\\\
				                                              & 01
				                                              & -3.0
				                                              & Different acceleration $\rm [m/s^2]$
				                                              & 0.03, 0.05, 0.07, 0.09, 0.11
				                                              & \multirow{2}{12em}{0.01, 0.02, 0.04, 0.06, 0.08, 0.10, 0.12, 0.14}
				\\\\
				                                              & 02
				                                              & -12.0
				                                              & \multirow{2}{14em}{Mixing of different factory noise [factory noise index]}
				                                              & A, B
				                                              & C
				\\\\ \hline
				\multirow{6}{*}{Valve}
				                                              & 00
				                                              & 0.0
				                                              & \multirow{2}{14em}{Different open/close operation patterns [pattern index]}
				                                              & 00, 01
				                                              & 02, 03
				\\\\
				                                              & 01
				                                              & 0.0
				                                              & \multirow{2}{14em}{Different number and location of panels [panel locations]}
				                                              & \multirow{2}{12em}{open (no panels), bs-c (back-side closed)}                                 & \multirow{2}{12em}{b-c (back closed), s-c (side closed)}
				\\\\
				                                              & 02
				                                              & 0.0
				                                              & \multirow{3}{14em}{Different number of valves [(valve1 pattern index, valve2 pattern index)]}
				                                              & \multirow{2}{12em}{(v1 04), (v1 05), (v2 04), (v2 05)}                                        & \multirow{2}{12em}{(v1 04, v2 04), (v1 04, v2 05), (v1 05, v2 04), (v1 05, v2 05)}
				\\\\\\ \hline
			\end{tabular}\\
		\end{center}
	\end{table*}

	\begin{figure}[t]
		\begin{center}
			\begin{minipage}{0.49\hsize}
				\begin{center}
					\includegraphics[width=1.0\hsize,height=2.0cm,clip]{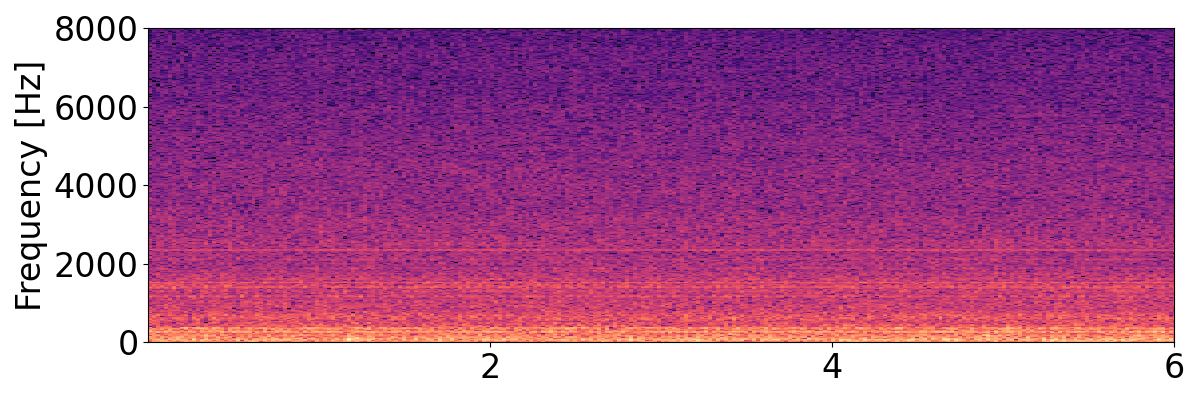}\\
					(a) Fan\\
				\end{center}
			\end{minipage}
			\begin{minipage}{0.49\hsize}
				\begin{center}
					\includegraphics[width=1.0\hsize,height=2.0cm,clip]{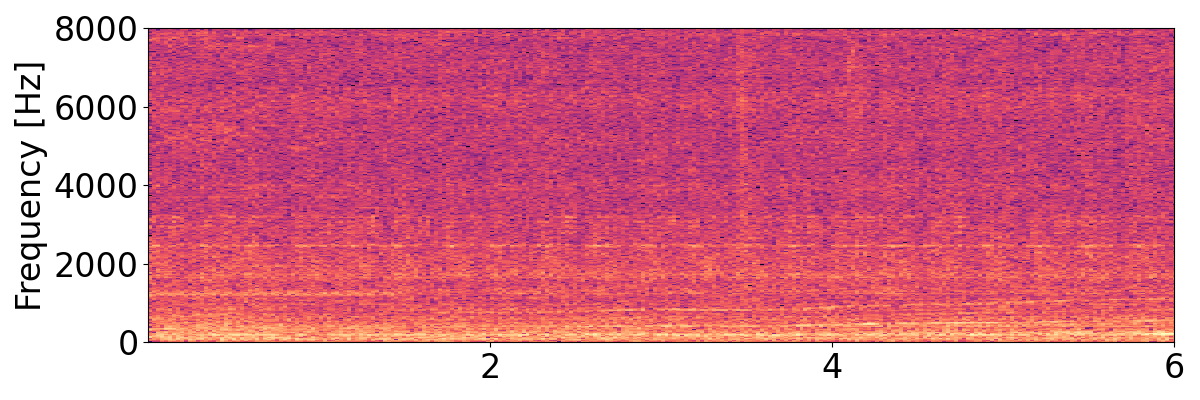}\\
					(b) Gearbox\\
				\end{center}
			\end{minipage}
			\begin{minipage}{0.49\hsize}
				\begin{center}
					\includegraphics[width=1.0\hsize,height=2.0cm,clip]{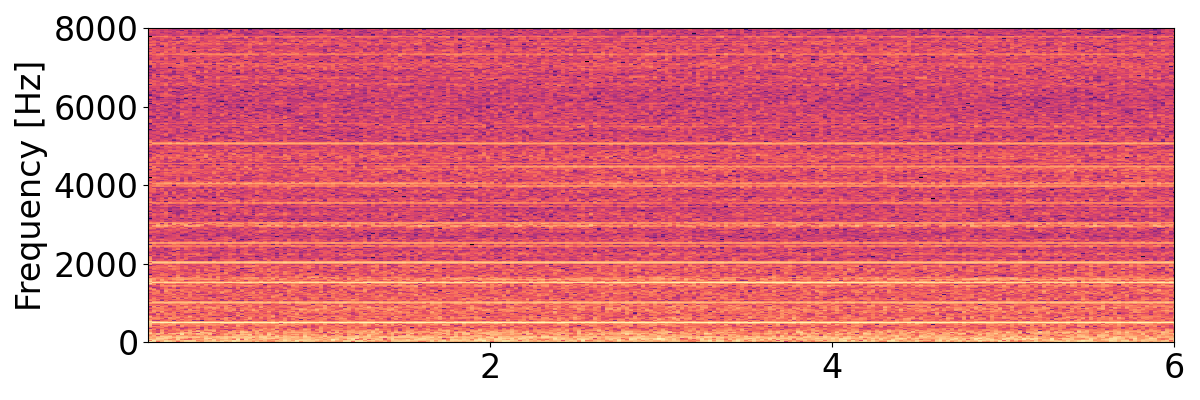}\\
					(c) Bearing\\
				\end{center}
			\end{minipage}
			\begin{minipage}{0.49\hsize}
				\begin{center}
					\includegraphics[width=1.0\hsize,height=2.0cm,clip]{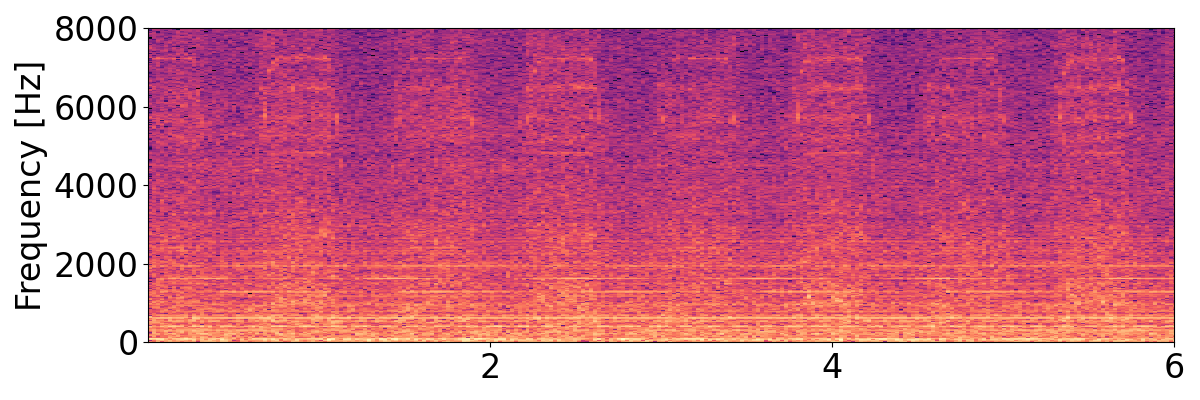}\\
					(d) Slider\\
				\end{center}
			\end{minipage}
			\begin{center}
				\includegraphics[width=0.49\hsize,height=2.0cm,clip]{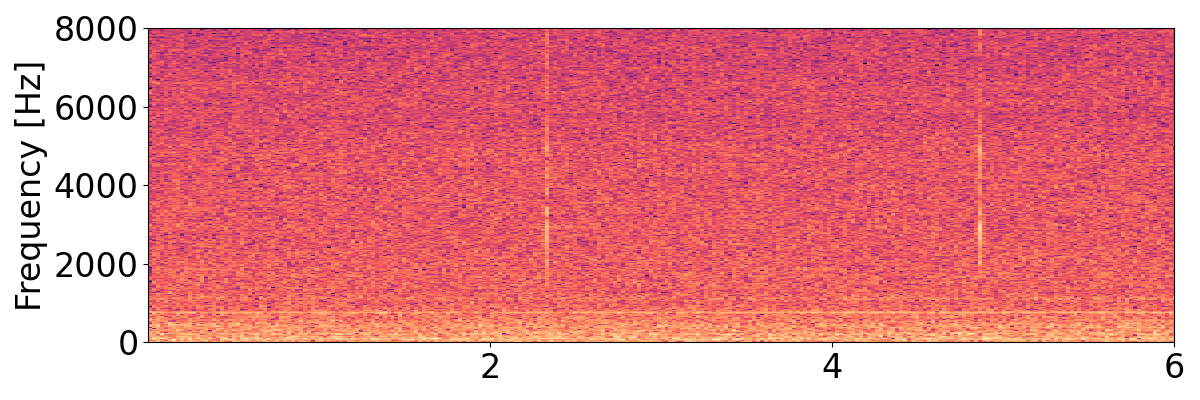}\\
				(d) Valve\\
			\end{center}
			\caption{Examples of spectrograms for each machine type.}
		\end{center}
		\vspace{-10pt}
		\label{fig:spec}
	\end{figure}


	After recording the machine sounds, we mixed the prerecorded factory noise A, B, or C as the background noise to simulate real-world environments. The factory noise A, B, and C were recorded in different real factories and consisted of sounds of various machinery. The noise-mixed data of each section was generated by the following steps.
	\begin{enumerate}
		\item The average power over all clips in the section, $a$ was calculated.
		\item For each clip $i$ from the section,
		      \begin{enumerate}
			      \item the signal-to-noise ratio (SNR) $\gamma$ dB for the clip was set to the value shown in Table \ref{parameters},
			      \item a background-noise clip $j$ was randomly selected, and its power $b_{j}$ was
			            tuned so that $\gamma = 10 \log_{10} (a/b_{j})$, and
			      \item the noise-mixed data was generated by mixing the machine sound clip $i$ and
			            the power-tuned background-noise clip $j$.
		      \end{enumerate}
	\end{enumerate}
	Here, the background-noise clip $j$ was randomly selected from predetermined types of factory noise, depending on the domain shift scenario. For
	Fan section 01, Bearing section 02, and Slide rail section 02, factory noise A and B were used for source domain and factory noise C was used for target domain. For other sections, factory noise A and B were used for both source and target domain.
	Also, for Fan section 00, we additionally mixed sound data of pumps from MIMII DUE.


	The complete dataset consists of normal and
	anomalous operating sounds of five different types of industrial machines, and
	each machine type has three sections with source and target domain samples.
	Table \ref{content} lists the number of samples in each section.
	The training data have 990 source domain samples and ten target domain samples for each
	section. We prepared ten target domain samples for training data so that the users can utilize
	a small number of target samples for generalization if the generalization of the model was too difficult.
	The test data have 50 normal samples and 50 anomalous samples for both domains.

	\begin{table}
		\begin{center}
			\caption{Number of samples in each section}
			\label{content}
			\begin{tabular}{@{}l c c| c c@{}}
				      & \multicolumn{2}{c|}{Source domain} & \multicolumn{2}{c}{Target domain}                    \\ \cline{2-5}
				      & normal                             & anomaly                           & normal & anomaly \\ \hline
				Train & 990                                & 0                                 & 10     & 0       \\
				Test  & 50                                 & 50                                & 50     & 50      \\ \hline
			\end{tabular}\\
		\end{center}
	\end{table}

	\section{Relation to MIMII DUE and ToyADMOS2}
	\label{sec:format}
	While MIMII DUE and ToyADMOS2 were developed for domain adaptation tasks, MIMII DG in this paper is for domain generalization tasks.
	As described in Sec. \ref{sec:intro}, the domain generalization techniques are promised for handling domain shifts that domain adaptation techniques may not be applicable. We created a new dataset dedicated to the domain generalization tasks because the dataset for domain generalization tasks and domain adaptation tasks should be different in some points.
	We included these points as three main features that characterize differences from MIMII DUE and ToyADMOS2.


	\begin{itemize}
		\item The number of values the domain shift parameter (a parameter that causes domain shift) takes has increased to at least three for each type of domain shift. This change is crucial because domain generalization techniques may require multiple sets of data obtained from different domain shift parameter values to generalize the model \cite{Dohi2021}. For example, for the velocity shift in Slide rail, we increased the number of values of the velocity from four in MIMII DUE to 13 in MIMII DG. Also, with the increased number of sets, users can adjust the difficulty of the generalization task.
		\item Domain shifts that can be difficult to detect are introduced. As described in Sec. \ref{sec:intro}, domain generalization techniques are preferred for domain shifts that can be unnoticed. Therefore, we introduced difficult-to-detect domain shifts such as differences in states of a machine operating in the background.
		\item Domain shift parameters become easier to access and utilize. To generalize the model, not only the sound data but additional information such as the domain shift parameters and other attributes can be useful. Therefore, easy access to these additional information is crucial. Unlike MIMII DUE, we specified domain shift parameters in file names and attribute files for both the source and target domain. With domain shift parameters in the target domain, users can evaluate the detection performance for each value of the domain shift parameter.
	\end{itemize}

	\section{Experiment}
	\label{sec:majhead}
	In this section, we use MIMII DG to benchmark the domain generalization performance of two baseline systems.


	\subsection{Baseline systems}
	We used two ASD systems for benchmarking; an autoencoder-based system and a MobileNetV2-based
	system.
	These systems are provided as the baseline systems in
	Task 2 of the DCASE 2022 Challenge, and Python implementations of the systems
	are available at \url{https://github.com/Kota-Dohi/dcase2022_task2_baseline_ae}
	for the autoencoder-based system and
	\url{https://github.com/Kota-Dohi/dcase2022_task2_baseline_mobile_net_v2}
	for the MobileNetV2-based system.

	The autoencoder-based system is often used as an unsupervised ASD system.
	Sound data were first converted to log-Mel spectrogram with a frame size of 1024,
	a hop size of 512, and 128 Mel bins. Five frames with four overlappings were
	successively concatenated to generate
	640-dimensional input feature vectors. The model had four linear layers with
	128 dimensions for the encoder, one bottle-neck layer with eight dimensions,
	and four linear layers with 128 dimensions for the decoder.
	The model was trained to minimize the
	error between the input feature vector $\mathbf{x}$ and the reconstruction $\mathbf{x'}$.
	We trained the model for 100 epochs using the Adam optimizer\cite{Kingma2014} with a learning rate of
	0.0001 and a batch size of 128.
	The anomaly scores were calculated by the averaged reconstruction error.

	The MobileNetV2-based system uses an auxiliary task to improve the detection performance
	of an unsupervised ASD system \cite{giri2020self,primus2020anomalous}.
	64 frames with 48 overlappings were successively concatenated to generate
	input feature vectors.
	For the model, we used a MobileNetV2 \cite{Sandler2018} with a multiplier parameter of 0.5.
	The model was trained to classify section IDs for each machine type.
	We trained the model for 20 epochs using the Adam optimizer with a learning rate of
	0.0001 and a batch size of 128.
	The anomaly scores were calculated by the averaged negative logit of the
	predicted probabilities for the correct section.

	\subsection{Metric}
	We used the area under the receiver operating characteristic curve (AUC) for evaluation.
	Because the domain generalization task requires detecting anomalies even when
	the occurance of domain shifts can be difficult to detect, the anomaly detector is expected to work with the same threshold regardless of the domain.
	Therefore, we calculated the AUC using both the source and target domain data.
	Also, to evaluate the anomaly detection performance for each domain, the AUC was computed for each domain. The AUC for each domain, section, and machine type was calculated as
	\begin{equation}
		{\rm AUC} = \frac{1}{N^{-}_{d}N^{+}_{n}} \sum_{i=1}^{N^{-}_{d}} \sum_{j=1}^{N^{+}_{n}}
		\mathcal{H} (\mathcal{A}_{\theta} (x_{j}^{+}) - \mathcal{A}_{\theta} (x_{i}^{-})),
	\end{equation}
	where $n$ represents the index of a section,
	$d \in \{ {\rm source}, {\rm target} \}$ represents a domain,
	and $\mathcal{H} (x)$ returns 1 when $x > 0$ and 0 otherwise.
	$\mathcal{A}_{\theta}(x)$ is the anomalous score of a sound clip $x$, where $\theta$ is the parameters of the system.
	Here, $\{x^{-}_{i}\}_{i=1}^{N^{-}_{d}}$ is normal test clips in the domain $d$ in the section $n$
	and $\{x_{j}^{+}\}_{j=1}^{N^{+}_{n}}$ is anomalous test clips in the section $n$ in the machine type $m$.
	$N^{-}_{d}$ is the number of normal test clips in the domain $d$
	and $N^{+}_{n}$ is the number of anomalous test clips in the section $n$.

	\begin{table}[t]
		\begin{center}
			\caption{AUC (\%) of each domain for each section.}
			\label{results}
			\begin{tabular}{@{}l l r r r r@{}}
				\hline
				                                               &           &
				\multicolumn{2}{c}{\begin{tabular}{@{\hskip0pt}l@{\hskip0pt}} \textbf{ Autoencoder } \end{tabular}} &
				\multicolumn{2}{c}{\begin{tabular}{@{\hskip0pt}l@{\hskip0pt}} \textbf{ MobileNetV2 } \end{tabular}}                                                           \\ \cline{3-6}
				\multicolumn{2}{c}{\begin{tabular}{@{\hskip0pt}l@{\hskip0pt}} \textbf{Machine type / }\\ \textbf{section ID} \end{tabular}} &
				\begin{tabular}{@{\hskip0pt}r@{\hskip0pt}}  \textbf{source}\\  \end{tabular}                     &
				\begin{tabular}{@{\hskip0pt}r@{\hskip0pt}}  \textbf{target}\\ \end{tabular}                     &
				\begin{tabular}{@{\hskip0pt}r@{\hskip0pt}}  \textbf{source}\\  \end{tabular}                     &
				\begin{tabular}{@{\hskip0pt}r@{\hskip0pt}}  \textbf{target}\\ \end{tabular}                                                                               \\
				\hline
				\multirow{3}{*}{Fan}
				                                               & 00        & $84.69$   & $39.35$   & $71.07$   & $62.13$ \\
				                                               & 01        & $71.69$   & $44.74$   & $76.26$   & $35.12$ \\
				                                               & 02        & $80.54$   & $63.49$   & $67.29$   & $58.02$ \\ \hline
				\multirow{3}{*}{Gearbox}
				                                               & 00        & $64.63$   & $64.79$   & $63.54$   & $67.02$ \\
				                                               & 01        & $67.66$   & $58.12$   & $66.68$   & $66.96$ \\
				                                               & 02        & $75.38$   & $65.57$   & $80.87$   & $43.15$ \\ \hline
				\multirow{3}{*}{Bearing}
				                                               & 00        & $57.48$   & $63.07$   & $67.85$   & $60.17$ \\
				                                               & 01        & $71.03$   & $61.04$   & $59.67$   & $64.65$ \\
				                                               & 02        & $42.34$   & $52.91$   & $61.71$   & $60.55$ \\ \hline
				\multirow{3}{*}{Slide rail}
				                                               & 00        & $81.92$   & $58.04$   & $87.15$   & $80.77$ \\
				                                               & 01        & $67.85$   & $50.30$   & $49.66$   & $32.07$ \\
				                                               & 02        & $86.66$   & $38.78$   & $72.70$   & $32.94$ \\ \hline
				\multirow{3}{*}{Valve}
				                                               & 00        & $54.24$   & $52.73$   & $75.26$   & $43.60$ \\
				                                               & 01        & $50.45$   & $53.01$   & $54.78$   & $60.43$ \\
				                                               & 02        & $51.56$   & $43.84$   & $76.26$   & $78.74$ \\ \hline
				\multicolumn{2}{c}{Average}                    & $ 67.21 $ & $ 53.99 $ & $ 68.72 $ & $ 56.42 $           \\ \hline
			\end{tabular}\\
		\end{center}
	\end{table}

	\subsection{Results}
	Baseline results are shown in Table \ref{results}.
	On average, the AUC for the target domain data was lower than the source domain data at 13.2\% with the autoencoder-based system and 12.3\% with the MobileNetV2-based system. In some sections, the AUC of the target domain was slightly higher than that of the source domain. This could be because the target domain data happened to be similar to the source domain data of other sections.
	Overall, the fact that models trained with the source-domain tended to show lower performance for the target data indicate that there is a significant difference between the source-domain data and the target-domain data. This suggests that domain shift scenarios have been successfully reproduced.
	Thus, the dataset is useful for benchmarking the performance of domain generalization techniques.

	\section{Conclusion}
	We presented a new dataset, MIMII DG, which  was developed for benchmarking domain generalization techniques for ASD.
	The dataset has normal and anomalous operating sounds of five different types of industrial machines with domain shifts.
	Experimental results using two ASD systems demonstrate that the detection performance significantly degrades for the target domain.

	\bibliographystyle{IEEEtran}
	\bibliography{refs}

\end{sloppy}
\end{document}